\documentclass[twocolumn,aps,showpacs,floatfix,prc]{revtex4}
\usepackage[dvips]{epsfig}
\begin{document}

\title{Alpha decay chains from element 113 }

\author{P. Roy Chowdhury$^1$ , D.N. Basu$^2$, C. Samanta$^{1,3}$ }
\affiliation{ $^1$ Saha Institute of Nuclear Physics, 1/AF Bidhan Nagar, Kolkata 700 064, India }
\affiliation{ $^2$ Variable  Energy  Cyclotron  Centre, 1/AF Bidhan Nagar, Kolkata 700 064, India }
\affiliation{ $^3$ Physics Department, Virginia Commonwealth University, Richmond, VA 23284-2000, U.S.A.}

\email[E-mail: ]{partha.roychowdhury@saha.ac.in} 
\email[E-mail: ]{dnb@veccal.ernet.in}
\email[E-mail: ]{chhanda.samanta@saha.ac.in}

\date{\today }

\begin{abstract}

Theoretical estimates of $\alpha$-decay half lives of several nuclei in the decay from element 113 are presented. Calculations in a WKB framework using DDM3Y interaction and experimental $Q$-values are in good agreement with the experimental data. Half life calculations are found to be quite sensitive to the $Q$-values and angular momenta transfers. Calculated decay lifetime decreases, owing to more penetrability as well as thinner barrier, as $Q$-value increases. Deviations to this predominant behaviour observed in some recent experimental data may be attributed to non zero spin-parities in some cases.
  
\end{abstract}


\maketitle

      In past two decades, syntheses of superheavy elements and their decay lifetime measurements have been two major goals of nuclear physics. This field has got new impetus by recent syntheses of several heavy elements and their subsequent re-confirmations~\cite{ho00,oga04,og04,mo04,oga06}. With the advent of radioactive ion beam facilities it is now believed that ultimately it would be possible to reach the center of the island of superheavy elements. In this scenario, study of decay properties of superheavy elements, primarily by $\alpha$ emissions, has become an important domain of intense research. Discoveries of new superheavy elements have also provided a testing ground for many theoretical formalisms. It has been shown that half life calculations in the WKB framework for barrier penetration using the density dependent effective nuclear interactions and experimental $Q$-values can well reproduce the experimental data on superheavy nuclei~\cite{prc06}. In a subsequent calculation, following the same WKB framework, it was shown~\cite{Zh06} that the Generalized Liquid Drop Model (GLDM) including the proximity effects (between nucleons in the neck or the gap between nascent fragments) could reasonably estimate the experimental data on superheavy nuclei when the experimental $Q$-values were used.

      In this work, half lives of parent nuclei $^AZ$, with the charge number Z~=113-107  decaying via $\alpha$ emissions are calculated in the WKB barrier penetration framework using microscopic potentials for the $\alpha$ - nucleus interaction. The nuclear potentials have been obtained microscopically by double folding the $\alpha$ and daughter nuclei density distributions with the density dependent M3Y (DDM3Y) effective nucleon-nucleon interaction. This procedure of obtaining nuclear interaction energy for the $\alpha$ - nucleus interaction is more fundamental in nature. Moreover, the use of a microscopic nuclear potential for a wide range of $\alpha$ - nucleus interaction~\cite{Ba03} is also a profound theoretical approach. The double folding potential, thus obtained, has been utilised for calculating the barrier penetration probability. The barrier pentrability is then used to provide estimates of $\alpha$ decay half lives for Z~=113-107 $\alpha$-emitters.  

      The decay $Q_{ex}$ values for the favored decays have been calculated from the measured $\alpha$ particle kinetic energies $E_{\alpha}$ using standard recoil correction and the electron shielding correction in a systematic manner as suggested by Perlman and Rasmussen\cite{Pe57}. The decay $Q_{ex}$ value and the  measured $\alpha$ particle kinetic energy $E_{\alpha}$ are related by the following expression: 

\begin{equation}
 Q_{ex} = (\frac{A_p}{A_p-4})E_{\alpha} + (65.3 Z_p^{7/5} - 80.0 Z_p^{2/5}) \times 10^{-6} ~\rm MeV
\label{seqn1}
\end{equation}
\noindent
where the first term in the right hand side is the standard recoil correction and the second term is an electron shielding correction. 

      Table-I shows that the calculated values are in good agreement with most of the experimental data~\cite{oga04,mo04}, but it also indicates some possible discrepancies in some experimental data on the $\alpha$-decay half lives of $^{278}113, ^{274}111, ^{270}109$ and, $^{266}107$~\cite{mo04}. Owing to more penetrability as well as thinner barrier in case of higher $Q$ value, the $T_{1/2}$ value decreases as $Q$ increases. However, the decay times of the ref.~\cite{mo04} show opposite trend for $^{274}111$ and $^{270}109$ nuclei. Since these nuclei are all odd-odd nuclei, they have non-zero spins. In some cases, the spin-parity conservation might force an $\alpha$-particle to carry away non-zero angular momentum. The additional barrier arising due to the centrifugal contribution $\hbar^2 l(l+1) / (2\mu R^2)$, where $R$ and $\mu$ are the distance and reduced mass of the $\alpha$-daughter system respectively, acts to reduce the tunneling probability if the angular momentum $l$ carried away by the $\alpha$-particle is non-zero. Hindrance factor which is defined as the ratio of the experimental $T_{1/2}$ to the theoretical $T_{1/2}$ is therefore larger than unity since the decay involving a change in angular momentum can be strongly hindered by the centrifugal barrier. 

      The measured decay time, according to Ref.~\cite{mo04}, is only for one observed event. In case of better statistics this corresponds to mean lifetime for the decay. The half lives are just 0.693 (= ln2) times the mean lives. In the low statistics measurement, the half lives can be estimated approximately by considering the measured decay time value for only one atom measurement~\cite{mo04} as the mean life. These values however do not agree well with the theoretical calculations for $l=0$ angular momenta transfers to the emitted $\alpha$-particles. Moreover, the calculated decay lifetimes should decrease as $Q$-values increase. As stated before, deviations to this  predominant behaviour is observed in these experimental data. This might be due to non zero spin-parities of these odd-odd nuclei and the spin-parity conservation might force the emitted $\alpha$-particles to carry non-zero $l$.

      In case of $^{278}113$ for a centrifugal barrier arising out of $l=3$ angular momentum transfer to an $\alpha$-particle, using experimental $Q$-value ($Q_{ex}$), the theoretical half life becomes $307^{+67}_{-55}$$\mu$s and in case of $^{274}111$ half life becomes $6.45^{+2.83}_{-1.98}$ms corresponding to $l=5$. These can provide good explanations for the measured lifetimes. Consideration of angular momenta transfers of 2 and 4 units for $^{278}113$ and $^{274}111$, respectively, instead of 3 and 5, gives theoretical half lives as $176^{+38}_{-31}$$\mu$s and $2.57^{+1.14}_{-0.78}$ms respectively. Experimentally measured half lives corresponding to $^{270}109$ and $^{266}107$, which are lower than the present esimates, warrant further experimental measurements with higher statistics. Theoretical decay $Q$-values $Q_{th}$ from Refs.~\cite{MU03,Mu03} which provide best theoretical estimates in this mass domain are also presented in Table-I alongwith the results of the present calculations for half-lives using these theoretical $Q$-values.

\begin{table*}[h]
\caption{\label{table1}Comparison between experimental and calculated $\alpha$-decay half-lives of nuclei for zero angular momenta transfers using experimental and theoretical [10,11] $Q$-values. } 
\begin{ruledtabular}
\begin{tabular}{lllllll}
 Parent &Expt.&Theo. &Expt.& DDM3Y&DDM3Y&Refs.   \\ 
 Nuclei&$(MeV)$&$(MeV)$&&This work&This Work&\\  
\hline
$^A Z$&$Q_{ex}$&$Q_{th}$&$T_{1/2}$&$T_{1/2}[Q_{ex}]$&$T_{1/2}[Q_{th}]$&Expt.\\ \hline
&&&&\\
$^{284}113$&$10.15(6)$&10.68&$0.48^{+0.58}_{-0.17} $s&$1.55^{+0.72}_{-0.48} $s&0.06 s&\cite{oga04} \\ 
&&&&\\
$^{283}113$&$10.26(9)$&11.12&$100^{+490}_{-45} $ms&$201.6^{+164.9}_{-84.7} $ms&1.39 ms&\cite{oga04} \\ 
&&&&\\
$^{278}113$&$11.90(4)$~$^{a)}$&-------&344 $\mu$s~$^{b)}$&$101^{+27}_{-18}$ $\mu$s&-------&\cite{mo04}  \\ 
&&&&\\
$^{280}111$&$9.87(6)$&10.77&$3.6^{+4.3}_{-1.3} $s&$1.9^{+0.9}_{-0.6} $s&0.01 s&\cite{oga04}  \\ 
&&&&\\
$^{279}111$&$10.52(16)$&11.08&$170^{+810}_{-80} $ms&$9.6^{+14.8}_{-5.7} $ms&0.42 ms&\cite{oga04}  \\ 
&&&&\\
$^{274}111$&$11.36(7)$~$^{a)}$&11.53& 9.26 ms~$^{b)}$&$0.39^{+0.18}_{-0.12} $ms&0.17ms&\cite{mo04}  \\ 
&&&&\\
$^{276}109$&$9.85(6)$&10.09&$0.72^{+0.87}_{-0.25} $s&$0.45^{+0.23}_{-0.14} $s&0.10 s&\cite{oga04}  \\ 
&&&&\\
$^{275}109$&$10.48(9)$&10.34&$9.7^{+46}_{-4.4} $ms&$2.75^{+1.85}_{-1.09} $ms&6.36 ms&\cite{oga04}  \\ 
&&&&\\
$^{270}109$&$10.23(7)$~$^{a)}$&10.27&7.16 ms~$^{b)}$&$52.05^{+27.02}_{-17.68} $ms&41.10 ms&\cite{mo04}   \\ 
&&&&\\
$^{272}107$&$9.15(6)$&9.08&$9.8^{+11.7}_{-3.5} $s&$10.1^{+5.4}_{-3.4} $s&16.8 s&\cite{oga04}  \\ 
&&&&\\
$^{266}107$&$9.26(4)$~$^{a)}$&8.95&2.47 s~$^{b)}$&$5.73^{+1.82}_{-1.38} $s&50.83 s&\cite{mo04} \\ 
\end{tabular} 
\end{ruledtabular}
$a)$ These  $Q_{ex}$ values are calculated using the measured $\alpha$-decay energies~\cite{mo04}. \\
$b)$ These are experimental decay times~\cite{mo04} (not $T_{1/2}$ values). 
\end{table*}

      Interestingly some discrepancies are reported in decay times in a recent repeat experiment in which (from two events) the mean life has been deduced by averaging two values~\cite{mo07}. The measured values along with corresponding theoretical calculations are listed in Table-II. Although the $Q_{ex}$ value is slightly lower for $^{274}111$ than that for $^{278}113$, still the $T_{1/2}[Q_{ex}]$ value for $^{274}111$ is slightly lower than that for $^{278}113$,  due to higher Coulomb and nuclear potentials for the latter case \cite{prc06}. The comparison suggests rather high angular momenta transfers to the emitted $\alpha$-particles. In this regard, caution should be exercised in making any specific conclusion for these four nuclei (as the experimental data are from only two events) before further experimental measurements with higher statistics are available.

\begin{table*}[h]
\caption{\label{table2}Comparison between new experimental [12] and calculated $\alpha$-decay half-lives of nuclei for 0 and 5 units of angular momenta transfers using experimental [12] and theoretical [10, 11] $Q$-values. Experimental $T_{1/2}$ = 0.693 $\times$ Assigned Mean-life of Ref.[12]. } 
\begin{ruledtabular}
\begin{tabular}{lllllll}
 Parent &Expt.&Theo. &Expt.& DDM3Y &DDM3Y &DDM3Y \\ 
 & & & &$l=0$ &$l=0$ &$l=5$ \\
 Nuclei&$(MeV)$&$(MeV)$&&This work&This work&This work \\  
\hline
$^A Z$&$Q_{ex}$&$Q_{th}$&$T_{1/2}$&$T_{1/2}[Q_{ex}]$&$T_{1/2}[Q_{th}]$&$T_{1/2}[Q_{ex}]$\\ \hline
&&&\\
$^{278}113$&$11.737(40)$&-------&$1.8^{+3.3}_{-1.3}$ ms&$0.23^{+0.05}_{-0.04}$ ms&-------& $3.71^{+0.68}_{-0.84}$ ms\\ 
&&&\\
$^{274}111$&$11.525(71)$&11.53&$15.2^{+27.7}_{-11.1} $ ms&$0.17^{+0.07}_{-0.05}$ ms &0.17 ms&$2.75^{+1.20}_{-0.84}$ ms\\ 
&&&\\
$^{270}109$&Escaped&10.27&$0.57^{+1.04}_{-0.41} $ s&-------&41.10 ms&$0.93^{+0.49}_{-0.32}$ s$^*$  \\ 
&&&\\
$^{266}107$&$9.964(41)$&8.95&$1.32^{+2.36}_{-0.97} $ s&$0.058^{+0.017}_{-0.013}$ s&50.83 s &$1.06^{+0.30}_{-0.24}$ s\\ 
\end{tabular} 
\end{ruledtabular}
$^*$ Using $Q_{ex}=10.23 (7)$~\cite{mo04}, same as in Table 1.\\
\end{table*}

      In summary, the WKB framework with DDM3Y interaction is known to providee good estimates of the experimental data when experimental $Q$-values are used~\cite{prc06}. We have calculated the $\alpha$-decay half lives of the recent data on  element 113~\cite{oga04,mo04,mo07}. Interestingly, although the $T_{1/2}$ values should decrease as $Q$ value increases, the decay times of the ref.~\cite{mo04} show opposite trend for $^{274}111$ and $^{270}109$ nuclei. This discrepancy does not exist in the subsequent repeat experiment~\cite{mo07}. Comparison with theoretical calculations indicates possible non-zero angular momenta transfers but warrants further experimental measurements with higher statistics.

\end{document}